\documentclass[lettersize,journal]{IEEEtran}
\usepackage{amsmath,amsfonts}
\usepackage{algorithmic}
\usepackage{algorithm}
\usepackage{array}
\usepackage[caption=false,font=normalsize,labelfont=sf,textfont=sf]{subfig}
\usepackage{textcomp}
\usepackage{stfloats}
\usepackage{url}
\usepackage{verbatim}
\usepackage{graphicx}
\usepackage{cite}
\usepackage{mathtools}
\usepackage{placeins}
\usepackage{float}
\usepackage{afterpage}
\usepackage[hidelinks]{hyperref}
\usepackage{xcolor}
\newcommand\hl[1]{%
  \bgroup
  \hskip0pt\color{red!80!black}%
  #1%
  \egroup
}
\begin{document}
%\title{Neural Networks Hear You Loud and Clear: A Sound Optimization Scheme for Hearing Loss Emulation}
\title{How to train your ears: Auditory-model emulation for large-dynamic-range inputs and mild-to-severe hearing losses}
%\iffalse
\author{Peter Leer, Jesper Jensen, Zheng-Hua Tan, ~\IEEEmembership{Senior Member, IEEE,} Jan Østergaard, ~\IEEEmembership{Senior Member, IEEE,} Lars Bramsløw% <-this % stops a space
        
\thanks{This work is partly supported by Innovation Fund Denmark Case no. 0153-00091B  }% <-this % stops a space
\thanks{The trained DNNs can be found at \url{https://github.com/P-Leer/HowToTrainYourEars} together with an example of usage.}%
\thanks{P. Leer is with Eriksholm Research Centre, 3070, Snekkersten, Denmark, and with the Department of Electronic Systems,
Aalborg University, 9220 Aalborg Øst, Denmark (e-mail: pelb@eriksholm.com).}%
\thanks{J. Jensen is with Demant A/S, 2750, Smørum, Denmark, and with the Department of Electronic Systems,
Aalborg University, 9220 Aalborg Øst, Denmark (e-mail: jesj@demant.com).}%
\thanks{J. Østergaard is with the Department of Electronic Systems,
Aalborg University, 9220 Aalborg Øst, Denmark (e-mail: jo@es.aau.dk).}%
\thanks{Z-H. Tan is with the Department of Electronic Systems,
Aalborg University, 9220 Aalborg Øst, Denmark Denmark, and also with the Pioneer Centre for
AI, 1350 Copenhagen, Denmark (e-mail: zt@es.aau.dk).}%
\thanks{L. Bramsløw is with Eriksholm Research Centre, 3070, Snekkersten, Denmark (e-mail: labw@eriksholm.com).}}%
%\fi
\markboth{}%
{Shell \MakeLowercase{\textit{et al.}}: Improved hearing loss emulation via deep neural networks and a novel optimization scheme}

\IEEEpubid{}

\maketitle

\begin{abstract}
%\boldmath
Advanced auditory models are useful in designing signal-processing algorithms for hearing-loss compensation or speech enhancement. Such auditory models provide rich and detailed descriptions of the auditory pathway, and might allow for individualization of signal-processing strategies, based on physiological measurements. However, these auditory models are often computationally demanding, requiring significant time to compute. To address this issue, previous studies have explored the use of deep neural networks to emulate auditory models and reduce inference time. While these deep neural networks offer impressive efficiency gains in terms of computational time, they may suffer from uneven emulation performance as a function of auditory-model frequency-channels and input sound pressure level, making them unsuitable for many tasks. In this study, we demonstrate that the conventional machine-learning optimization objective used in existing state-of-the-art methods is the primary source of this limitation. Specifically, the optimization objective fails to account for the frequency- and level-dependencies of the auditory model, caused by a large input dynamic range and different types of hearing losses emulated by the auditory model. To overcome this limitation, we propose a new optimization objective that explicitly embeds the frequency- and level-dependencies of the auditory model. Our results show that this new optimization objective significantly improves the emulation performance of deep neural networks across relevant input sound levels and auditory-model frequency channels, without increasing the computational load during inference. Addressing these limitations is essential for advancing the application of auditory models in signal-processing tasks, ensuring their efficacy in diverse scenarios.
\end{abstract}

% === KEYWORDS ====================================================================
% =================================================================================

 \IEEEpubidadjcol 
\section{Introduction}
\label{sec:introduction}
\begin{figure*}[!th]
  \centering
      \includegraphics[scale=0.6]{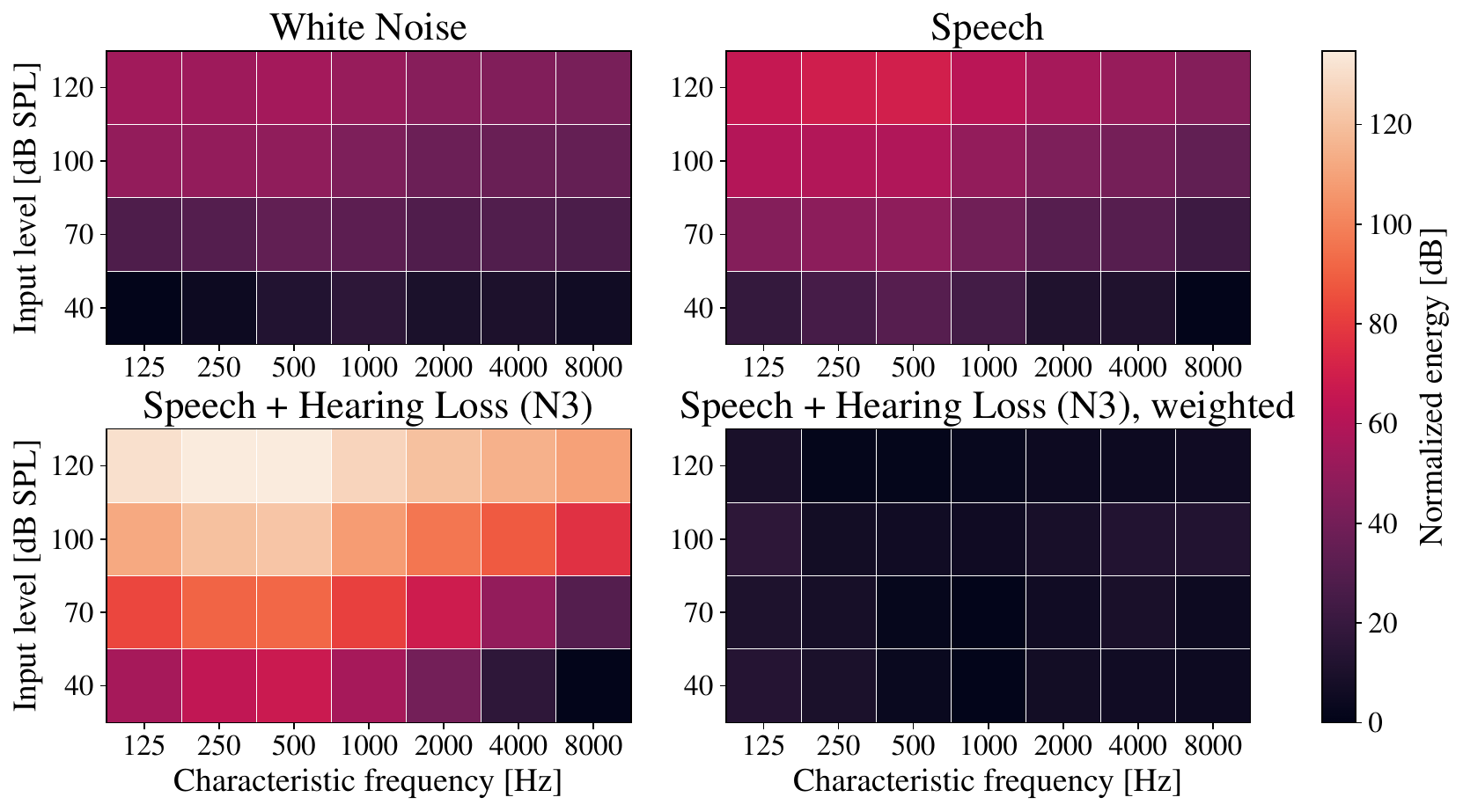}
  \caption{The energy distribution measured in the inner representation space of the Zilany auditory model \cite{Zilany2014}, as a function of characteristic frequency and input level for four different conditions: Normal hearing with white noise as input, normal hearing with speech as input, a N3 hearing loss with speech as input, a N3 hearing loss with speech as input and weighted according to the proposed scheme, cf. Sec \ref{sec:opt_and_eval}. The energies are calculated using 10 speech sentences and normalized relative to the lowest channel energy for each condition.}
  \label{fig:energy_distribution}
\end{figure*}

Auditory models have recently been applied for machine-learning tasks, such as designing speech- and audio-processing algorithms for hearing-assistive devices \cite{Wen}\cite{Drakopoulos2023ACompensation}. These auditory models represent various stages of the human auditory pathway, including the outer- and middle-ear, the cochlea and the auditory nerve. The auditory models are configured by a set of parameters allowing the user to model potential impairments, e.g. hair-cell loss, synaptopathy, etc. \cite{OssesVecchi2022}. Thus, auditory models can potentially enable development of novel and personalized signal processing strategies for hearing-assistive devices that take into account the dysfunction of the individual auditory pathway, as measured by physiological metrics. In practice, the computational load of these auditory models can be very high, thus limiting the feasibility of using the models as bio-inspired loss functions for deep-learning models, e.g. for use in hearing-loss compensation or noise reduction strategies. Previous efforts have addressed this problem by training a deep neural network (DNN) to emulate the auditory models \cite{Nagathil2021ComputationallyProcessing,AsbjrnLeerBysted,Baby2021AApplications,Nagathil2023WaveNet-basedModel}. However, as we demonstrate in this paper, the previous approaches, which - except for \cite{Nagathil2023WaveNet-basedModel} - rely on conventional machine-learning optimization objectives, applied directly on the auditory model output, such as the mean-square error (MSE) or the mean-absolute error (MAE), do not perform well across relevant input sound pressure levels (SPLs) and frequency channels, which are crucial for hearing-loss compensation (HLC) and noise reduction (NR). When developing such strategies, the auditory-model emulator should perform well across a large range of input SPLs, since any HLC strategy need to be able to process signals ranging from just-noticeable to almost-uncomfortably loud, i.e. covering several orders of magnitudes of input SPLs. However, when training an auditory-model emulator for the previously described applications, one might experience training pathologies, such as unpredictable high-frequency and level-dependent behaviour. If left unaddressed, the training pathologies lead to essentially non-functioning emulators, which will be reflected in the developed HLC and NR strategies. 
 \IEEEpubidadjcol
 We show that the training pathologies are caused by two factors: 1) A skewed distribution of energy across the frequency channels of the auditory model, due to:
 \begin{itemize}
    \item The low-pass characteristic of speech \cite{Byrne1994AnSpectra}.
    \item The band-pass characteristic of peripheral stages of the auditory models (i.e. middle ear and inner hair cells) \cite{Altoe2017Model-basedCurves} \cite{Zilany2009ADynamics}.
    \item The low-pass characteristic of common hearing losses \cite{Bisgaard2010}.
\end{itemize} and 2) a skewed distribution of energy in the frequency channels of the auditory model caused by a training set consisting of input-output pairs in a large dynamic range, e.g. (35 -105) dB SPL, as was used in \cite{Nagathil2021ComputationallyProcessing}.
In order to better illustrate these two factors, Fig. \ref{fig:energy_distribution} shows the energy distribution (denoted as inner representation) of the inner hair cells (IHCs) of the Zilany auditory model \cite{Zilany2014}  at different input levels and auditory-model frequency channels with a given characteristic frequency (CF) for different inputs, both for a normal hearing parameterization and a N3 hearing loss (a moderate hearing loss, cf. Sec. \ref{sec:method_zilany}). Clearly, there is a difference of more than 12 orders of magnitudes in energy between the lowest CF and highest CF (e.g. the 8 kHz CF at 40 dB SPL vs the 125 Hz CF at 120 dB SPL), when the combined effect of a speech spectrum and the moderate hearing loss (N3) is introduced. The conventional optimization schemes, used in \cite{Nagathil2021ComputationallyProcessing},\cite{Baby2021AApplications} and \cite{Drakopoulos2021ASynapses}, will favour the performance in the high-energy regions of the energy distribution, and a DNN trained to emulate this auditory model will have uneven performance across both input SPLs and CFs. 

To circumvent the problems of conventional optimization objectives, we propose a DNN training method that embeds the dynamic range of the auditory model, as a function of frequency channel and input SPL, into the optimization objective. We measure the performance of our optimization objective using the MAE and our proposed evaluation metric, CF-dependent signal-to-error ratio (SER) across a wide range of sound pressure levels on DNN emulations of two different auditory models: 1) a biophysical transmission-line model of the basilar membrane, which is a part of the Verhulst model \cite{Verhulst2018} and 2) the cochlea stage of a parallel filter-bank model of the auditory nerve, that includes impairment of the OHCs and IHCs in the cochlea (the Zilany model \cite{Zilany2014}). The Verhulst auditory model has previously been emulated using deep neural networks \cite{Baby2021AApplications}, but the Verhulst auditory model only allows for modelling up to mild hearing losses, which motivates the choice of the Zilany auditory model \cite{Zilany2014} as a second auditory model; the Zilany auditory model can model several non-linear phenomena of the mammalian auditory system and can emulate hearing losses, ranging from normal hearing up to severe hearing losses \cite{Zilany2006ModelingPeriphery}.

In Sec. \ref{sec:auditory_models} we introduce a notational framework and the two auditory models used in this work. In Sec. \ref{sec:opt_and_eval} we present our proposed optimization objective and performance-evaluation schemes. In Sec. \ref{sec:DNN} we introduce the DNN architectures used for auditory-model emulation, while Sec. \ref{sec:training} introduces our training paradigm and choice of hyperparameters. In Sec. \ref{sec:results} we present and discuss our results, comparing the conventional optimization schemes to our proposed scheme for speech and tonal inputs. Finally, in Sec. \ref{sec:conclusion} we present a conclusion to our work.

\section{Auditory models}
\label{sec:auditory_models}
In this section we will describe the two auditory models.
We define a generic notation that we will be using for both auditory models:
Define a signal space $X \subset R^T$ and an inner representation space $I \subset R^{J \times T}$. In this context, $T$ is the number of samples of the input signal, and $J$ is the number of frequency channels. 
We denote the auditory model as the function $f^{\theta}: X \rightarrow I$,  and $\theta$ as the free parameters of the auditory model. For easier reading, we will suppress the $\theta$ notation, except where necessary. To emphasize a single frequency channel of the auditory model, we use the notation $f_{j}$ for the $j$-th frequency channel of the inner representation.

\subsection{The Verhulst model}
The Verhulst model is a model of the auditory pathway, consisting of several stages, including a transmission-line model of the basilar membrane, an inner-hair cell stage an auditory-nerve stage and a midbrain stage. For this work we use the first stage of the Verhulst model, the transmission-line model of the basilar membrane (BM), modeled as a set of coupled harmonic oscillators \cite{Verhulst2018}. We use $J=201$ frequency channels to correspond with previous work in which 201 CFs were used \cite{Baby2021AApplications}. Hearing loss is simulated by parameterizing the instantaneous non-linearities, such that the CF amplitude response to a single tone is reduced by the specified loss, for example a uniform hearing loss of 20 dB is a reduction of the gain of all oscillators by 20 dB to a single tone at the CF of the oscillators. The model can only capture hearing loss that can be attributed to outer hair cells (OHCs), which is maximally 35 dB at every CF for this auditory model \cite{GitHubHearingTechnology/CoNNear_cochlea}. We use the two hearing losses: 1) Flat20, a uniform 20 dB hearing loss, and 2) Slope20\_5, a 5 dB hearing loss up to 1 kHz, followed by a sloping hearing loss with 20 dB hearing loss at the highest audiometric frequency, 8 kHz.
\subsection{The Zilany model}
\label{sec:method_zilany}
The Zilany model is a parallel filterbank model that models the auditory-nerve response to an input acoustical signal. The model consists of several stages, including a middle-ear stage, a combined cochlea stage, simulating both outer and inner hair cells (IHCs), and an auditory-nerve stage  \cite{Zilany2014}. For this work we use the middle-ear and the combined cochlea stage, since this is the main part of the model that changes as a hearing loss is introduced.
For this particular auditory model the parameter set can be found by using a function, fitaudiogram2, supplied with the code for the auditory model \cite{AuditoryCenter}, that takes a given (pure-tone) audiogram as an input and produces the parameters for the OHCs and the IHCs for each CF. An audiogram provides a measure of hearing loss, relative to normal hearing, and is defined for 10 frequency bands and interpolated linearly on a log/dB scale for CFs between these frequency bands. For the results presented here, we use $J=32$ CFs. Note that the number of CFs chosen here does not necessarily reflect the number of CFs required for training a HLC- or NR-system.
To illustrate the performance of the proposed optimization objective for a wide span of different hearing losses, we choose to use the standard template audiograms from \cite{Bisgaard2010}, and in particular N0, N3, N5 and S1, denoting, respectively, normal hearing, a moderate hearing loss, a severe hearing loss, and a  steeply-sloped hearing loss, cf. Table \ref{tab:audiograms}. For all audiograms we attribute 2/3 of the hearing loss to the OHCs and 1/3 to the IHCs, which is the default setting in the fitaudiogram2 function. The parameters are denoted as $C^{OHC}_j \in [0,1]$ and $C^{IHC}_j \in [0,1]$, i.e. two parameters for each CF, with 0 denoting complete dysfunction and 1 normal hearing.
\begin{table*}[]
\centering
\caption{Audiograms used for the Zilany model. This table expresses the hearing loss in dB for each frequency band for 3 different template hearing losses\cite{Bisgaard2010}}
\label{tab:audiograms}
\begin{tabular}{llllllllllll}
\hline
\textit{Frequency / Template} &  & 250 & 375  & 500 & 750  & 1000 & 1500 & 2000 & 3000 & 4000 & 6000 \\ \hline
\textbf{N3}               &  & 35  & 35   & 35  & 35   & 40   & 45   & 50   & 55   & 60   & 65   \\
\textbf{N5}               &  & 65  & 67.5 & 70  & 72.5 & 75   & 80   & 80   & 80   & 80   & 80   \\
\textbf{S1}               &  & 10  & 10   & 10  & 10   & 10   & 10   & 15   & 30   & 55   & 70  \\ \hline 
\end{tabular}
\end{table*}

\section{Proposed optimization and evaluation criteria}
\label{sec:opt_and_eval}
\begin{figure}[bh!]
  \centering
      \includegraphics[scale=0.4]{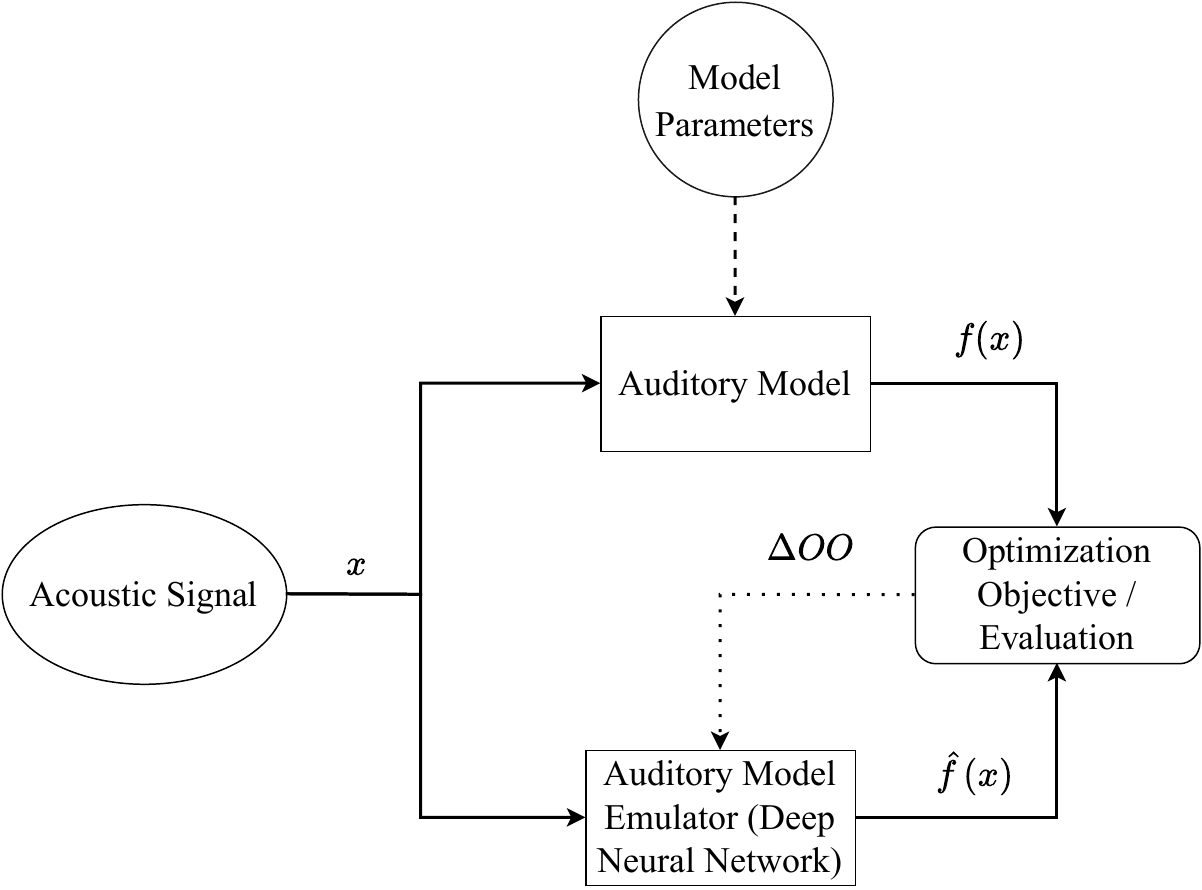}
  \caption{An overview of the auditory model emulator framework used in this work. The input signal is denoted by $x$, the reference auditory model output as $f(x)$, the output of the emulator as $\hat{f}(x)$ and the derivative of the optimization objective as $\Delta OO$. Bold lines denote the signal path, dashed lines denote the parameters, and the dotted line denotes the backpropagation of the Optimization Objective, which is used to train the Deep Neural Network.}
  \label{fig:overview} 
\end{figure}

In general, DNN emulators of auditory models are trained to minimize the difference between the inner representation of a ground-truth auditory model and an approximation by the DNN. The optimization objective measures this difference between the ground truth and the approximation by the DNN in the inner representation. This setup is shown in Fig. \ref{fig:overview}. In previous work, the mean-absolute error (MAE) and the mean-squared error (MSE) between the inner representations have been used as an optimization objective \cite{Nagathil2021ComputationallyProcessing,  Baby2021AApplications}.

The MAE averages the absolute error between the auditory model reference at a given CF, $f_j(x)$, and the DNN emulation, $\hat{f}_j(x)$,  over all the CFs,
\begin{equation}
    \mathrm{MAE}(f(x),\hat{f}(x)) = \dfrac{1}{T J
    }\sum_{j=1}^J||f_{j}(x) -\hat{f}_{j}(x)||_1 \, ,
    \label{eq:MAE}
\end{equation}
whereas the MSE is given by:
\begin{equation}
    \mathrm{MSE}(f(x),\hat{f}(x)) =\dfrac{1}{T J
    } \sum_{j=1}^J||f_j(x)-\hat{f}_j(x)||_2^2 \, .
\end{equation}
If the energy ($||f_j(x)||_2^2$) was uniformly distributed in the inner representation as a function of CF and input SPL, the relative performance of the emulator would be approximately equal across CF and input SPL. However, as shown in Fig. \ref{fig:energy_distribution}, there is a range of up to 12 orders of magnitudes in energy between the low and high-energy regions. Thus, one should expect an auditory model emulator trained with MAE or MSE as an optimization objective to perform poorly in the low-energy regions, because the corresponding error would be negligible compared to the high-energy regions, an expectation that we confirm in Sec. \ref{sec:results} for the auditory model emulators proposed in \cite{Baby2021AApplications,AsbjrnLeerBysted}.

\subsection{Constructing an optimizaton objective}
\noindent  In order to counteract the highly skewed frequency-and-level energy distribution of the inner representation, one approach would be to normalize (whiten) all the auditory-model-output targets, $f_j(x)$, across CFs and input SPL, resulting in an inner representation energy distribution shown in Figure \ref{fig:energy_distribution} (lower, right). However, in order to use the model for inference on unseen signals, this normalization would require constructing a non-linear and invertible function that maps the inner representation from a SPL-and-frequency-normalized inner representation (the neural network output), to the un-normalized representation (the estimated auditory-model output), as a function of the input. A similar parametric function, that does not normalize but compresses the inner representation by applying a symmetric, logarithmic function, controlled by a compression parameter $d$, is proposed in \cite{Nagathil2023WaveNet-basedModel}.
Instead of constructing such an explicit, normalizing function, we modify the MAE objective (\ref{eq:MAE}), to include a pre-determined, data-dependent factor ($\beta_j$) that balances the inner representations at each frequency channel together with a pre-determined, data-dependent factor  ($\alpha_{j,l}$) that balances the MAE across input SPL and frequency channels, where $l$ denotes the input SPL and $j$ the frequency channel. These two factors could be combined into a single factor that normalizes the output of the inner representation. However, due to the level dependencies, one would have to calculate one such factor for each level, which - as mentioned before - would require construction of additional functions. Therefore, we split the single normalizing component into the two previously described components, such that if these two components are multiplied they result in the expected value of the reciprocal magnitude at each CF and input level.  We introduce a normalized neural-network output $\bar{f}_{j}(x)$ that estimates a CF-normalized inner representation $\beta_j f_j(x)$. Hence, our proposed optimization objective, Frequency-and-level-dependent Mean-Absolute Error (FMAE), is defined as\footnote{The FMAE can be generalized by changing the norm used in (\ref{eq:FMAE}), (\ref{eq:beta_unnorm}) and (\ref{eq:alpha_norm}).}:

\begin{equation}
\resizebox{0.5\textwidth}{!}{$
    \mathrm{FMAE}(f(x_l),\bar{f}(x_l); \beta_j,\alpha_{j,l}) \stackrel{\text{def}}{=} \dfrac{1}{T J
    } \sum_{j=1}^J||\beta_j f_{j}(x_l) -\bar{f}_{j}(x_l)||_1 \alpha_{j,l} \, . $}
    \label{eq:FMAE}
\end{equation}

Since $\bar{f}_{j}(x)$ estimates $\beta_j{f}_j(x)$, it follows that an estimator of $f_j(x)$ can be found as:
\begin{equation}
    \hat{f}_{j}(x) = \dfrac{\bar{f}_{j}(x)  }{\beta_j} \, .
    \label{eq:f_estimate}
\end{equation}

\subsubsection{Estimating $\beta_j$ and $\alpha_{j,l}$}
Consider a training set of unnormalized  signals, $S$. Generate a training set $X_l$ by sampling from S, and normalize the samples to a given SPL ($l$). Repeat this procedure for a discrete number of SPLs, $l_{\textrm{min}} \leq l \leq l_{\textrm{max}}$ and denote the union of all $X_l$ as $X$. Denote the set containing all SPL levels as $L$. Then: 
\begin{equation}X_l =\{x \in X \, \big| \, \, ||x||_2 = p_0 10^{l/20}\} \, , \end{equation} where $p_0$ is the reference sound pressure of 20 $\mu$Pa. Using $X$, we find candidates for $\beta_j$ and $\alpha_{j,l}$ as follows:

First, we construct $\beta_j$, the factor in (\ref{eq:FMAE}) that counteracts the average energy frequency distribution. To do so, let $\bar{\beta}_{j,l}$ denote, for each SPL, the average of the inverse of the $L^{1}$-norm:
\begin{equation}
    \bar{\beta}_{j,l}= \dfrac{1}{|X_l|} \sum_{x_l \in X_l} \dfrac{1}{||f_{j}(x_l)||_1}\, \label{eq:beta_unnorm} . \\
\end{equation}
Next, to find $\beta_j$, $\bar{\beta}_{j,l}$ is normalized at each level and averaged, making it independent of the input SPL:
\begin{equation}
    \beta_j = \dfrac{1}{|L|}\sum_{l\in L}\dfrac{\bar{\beta}_{j,l}}{  \text{min}_{j \in [1,J]}\, \bar{\beta}_{j,l}}\, .
\end{equation}
Second, we construct $\alpha_{j,l}$. Since $\beta_j$ accounts for the average frequency distribution, $\alpha_{j,l}$ should account for the remaining level-and-frequency dependencies. Let $\bar{\alpha}_{j,l}$ denote, for each SPL, the  inverse of the $L^1$-norm divided by $\beta_j$ at each input SPL in the training set:
\begin{align}
    \bar{\alpha}_{j,l} &= \dfrac{1}{\beta_j|X_l|}\sum_{x_l \in X_l}  \dfrac{1}{||f_{j}(x_l)||_1} \,  \label{eq:alpha_norm}.
\end{align}
In order to use the same learning rates when using the MAE and the proposed FMAE, we make sure that the gradients of the MAE and FMAE are similar at the highest input SPL used for training. To achieve this,  $\bar{\alpha}_{j,l}$ is normalized such that the average multiplicative factor in (\ref{eq:FMAE}) is unity at the highest input SPL:
\begin{align}
    \alpha_{j,l} &= \dfrac{J \bar{\alpha}_{j,l}}{\sum_{j=1}^J\bar{\alpha}_{j,l_{\mathrm{max}}}}\, .
\end{align}
However, since $\alpha_{j,l}$ can only be found for the discrete levels contained in $L$, the FMAE (\ref{eq:FMAE}) might need additional adjustments, depending on the DNN training procedure. 
\subsubsection{Training on windowed segments}
 Consider training a DNN-emulator, using windowed segments ($x^*$) of $x_l\in X_l$. The resulting segment ($x^*$) might originate from a relatively  high or low-level portion of the original signal, which means that the corresponding SPL value ($l^*$) might not be equal to $l$ and, hence, $\alpha_{j,l^*}$ can not be computed. This problem is solved by replacing $\alpha_{j,l}$ in (\ref{eq:FMAE}) with a function ($a_j(x)$), found by interpolating in the log-domain: Find $l^{-} =  \underset{l\in L \, | \, l \leq l^*}{\mathrm{argmin}}|l-l^*| $,  $l^{+} =  \underset{l \in L \, | \, l^*  \leq l}{\mathrm{argmin}}|l-l^*| $ and define $m=\dfrac{l^{+}-l^*}{l^{+}-l^{-}}$. Define a function $a_j(x^*)$ as:
\begin{equation}
    a_j(x^*) = \begin{cases} \alpha_{j,l_{\mathrm{min}}} & \text{if } l^*\leq l_{\mathrm{min}} \\ 
    \alpha_{j,l_{\mathrm{max}}} & \text{if }  l_{\mathrm{max}} \leq l^* \\
    10^{m \log_{10}(\alpha_{j,l^{-}})+(1-m)\log_{10} (\alpha_{j,l^{+}})} & \textrm{else}
    \, .
    \end{cases}
\end{equation}

\subsection{Proposed evaluation criterion}
\label{sec:eval}
As discussed in the previous section, the MAE (\ref{eq:MAE}) will be dominated by the high-energy regions of the inner representation, whereas contributions from the low-energy regions will be negligible.  This makes the MAE inadequate as an indicator of how well a DNN model can emulate an auditory model. Hence, we need to construct a metric that scales across different magnitudes of energy, which can be done by scaling the energy of the error signal in a single frequency channel relative to the energy of the target. Inspired by the conventional signal-to-noise ratio, we propose to use the signal-to-error ratio (SER), calculated at a set of input sound levels for each frequency channel, defined as:
\begin{equation}
\label{eq:SER}
    \mathrm{SER}(j) \stackrel{\text{def}}{=} 20 \log_{10} \dfrac{|| f_{j}(x) ||_2}{||f_{j}(x) -\hat{f}_{j}(x)||_2}\, .
\end{equation}
It should be noted that as the emulation of $f_j(x)$ improves, the SER increases. A similar metric was employed in \cite{Nagathil2023WaveNet-basedModel}. 

\section{Deep Neural network architectures}
\label{sec:DNN}
To show that the proposed optimization objective, FMAE (\ref{eq:FMAE}), works generally and does not depend on any particular DNN architecture, we apply it for emulating two different auditory models, using different architectures. In general, we opt to use convolutional U-net encoder-decoder structures, as they allow for modelling the long time-constants of the auditory system efficiently and at the same time can accommodate different input lengths, due to the convolutional nature of the network.
\subsection{Architecture of the DNN emulator for the Verhulst model}
In order to emulate the Verhulst auditory model, we use the CoNNear architecture from \cite{Baby2021AApplications}. This architecture was previously used to emulate the Verhulst auditory model, and consists of a convolutional encoder-decoder, where the encoder downsamples the input by successive strided convolutions, and the decoder upsamples the encoder representation by transposed convolutions. All layers have the same kernel-size and number of output channels. In order to be consistent with previous work, we use the exact same network parameters and configuration as in \cite{Baby2021AApplications}. The parameters can be found in Table \ref{tab:connear}.

\subsection{Architecture of the DNN emulator for the Zilany model}
In order to emulate the Zilany auditory model, we use a variation of the Wave-U-Net \cite{Stoller2018}, a convolutional encoder-decoder network, where the encoder downsamples the input by decimation, and the decoder upsamples the encoder representation by successive convolutions followed by interpolations. Before the decoder there is an embedding layer, which is a convolutional layer without downsampling. The network uses skip connections between the encoder and decoder, improving gradient flow and allowing the decoder to recover temporal information lost in the downsampling procedure. The hyperparameters were found by performing a non-exhaustive pilot-test, using different configurations. The two primary considerations when choosing the hyperparameters were, 1) the size of the receptive field (cf. Sec. \ref{sec:DNN_RF}) and the choice of activation functions (cf. Sec. \ref{sec:activations}).
The hyperparameters of the neural network can be found in Table \ref{tab:WUN_param}.

\subsection{Receptive field of the deep neural networks}
\label{sec:DNN_RF}
The receptive field (RF) of a DNN is the portion of input space that must be used to compute the output, similar to the number of taps of a FIR filter. Therefore, we can expect that the DNN emulator needs an RF at least as long as the longest impulse response in the auditory model. For auditory models in general, this occurs at the lowest CF at the lowest input levels. As the input SPL is increased, the transfer function of a given CF at the cochlea broadens, corresponding to a shorter impulse response \cite{Baker2006AuditoryMasking}.
The RF of the networks is given by :
\begin{equation}
    \label{eq:REF}
    RF = \sum_{n=1}^N\left((k_n-1)\prod_{i=1}^{n-1}d_i\right)+1 \, \text{[samples]} \, ,
\end{equation}
where $N$ is the number of layers in the encoder, $k_n$ is the kernel size of the $n$-th layer, and $d_i$ is the downsampling factor of the $i$-th layer, i.e the size of the stride or the decimation factor \cite{Araujo2019ComputingNetworks}. For the Verhulst auditory model, we do not get to choose the RF, since we will be using the CoNNear network\cite{Baby2021AApplications}, and its network parameters, cf. Table \ref{tab:connear}. As an example, an impulse response with the length of the RF of the CoNNear is at best enough to account for $93\%$ of the power of a 0.5s-long 60 dB SPL reference impulse response at a sampling rate of 20 kHz. This means that we should expect the emulator to have worse performance at low SPL inputs at low frequencies, compared to higher SPL inputs. One could easily achieve a larger RF by increasing the number of layers or the kernel size, but do not change the architecture in order to be consistent with the original CONNear model. Conducting the exact same experiment for the Zilany model, a RF of at least 800 samples is needed to account for $99.99\%$ of the energy of the impulse response input at the lowest CF (125 Hz). This is achieved with the parameters in Table \ref{tab:WUN_param}, and can be verified by substituting the DNN parameters into (\ref{eq:REF}).

\subsection{Choice of activation functions}
\label{sec:activations}
In \cite{Baby2021AApplications} and \cite{Drakopoulos2021ASynapses}, considerable effort is put into choosing the activation functions for auditory model emulators proposed there. The authors suggest that the activations should resemble the input-output relations of the auditory model and cross the origin to allow for positive and negative deflections of the basilar membrane, hence the authors choose the hyperbolic tangent (Tanh) for CoNNear. Along the same line of reasoning we found for the Zilany model that using a Tanh in the encoder allowed us to model the symmetric deflections around the origin of the basilar membrane, while using a PReLU activation in the decoder made it easier to model the rectification-like behaviour of the IHC.

\begin{table}[H]
\centering
      \caption{Network parameters of the CoNNear DNN.  N denotes the amount of Upsampling/Downsampling blocks.}
      \label{tab:connear}
\begin{tabular}{|l|l|}
\hline
N     &  4   \\ \hline
Kernel size                 & 64     \\ \hline
Depth & 128 \\ \hline
Encoder activation & Tanh  \\ \hline
Decoder activation & Tanh \\ \hline
Bias & False \\ \hline
\end{tabular}
\end{table}

\begin{table}[H]
\centering
      \caption{Network parameters of the Wave-U-Net DNN. N denotes the amount of Upsampling/Downsampling blocks.}
      \label{tab:WUN_param}
\begin{tabular}{|l|l|}
\hline
N     & 6   \\ \hline
Kernel size                 & 21     \\ \hline
Depth & 128 \\ \hline
Encoder activation & Tanh  \\ \hline
Decoder activation & PReLU \\ \hline
Bias & False \\ \hline

\end{tabular}
\end{table}

\section{Deep Neural Network Training}
\label{sec:training}
In order to train the DNN emulators, a dataset is created based on 2500 random utterances from the LibriTTS \cite{Zen2019LibriTTS:Text-to-Speech} database. The utterances (input) are scaled to have a random SPL between 40 and 120 dB SPL in 10 dB steps and fed through the auditory model (output). The input-output pairs are originally generated at 100 kHz and resampled using a sampling frequency of 20 kHz and segmented into windows of 2048 samples (102 ms). All models are trained for 300 epochs, using the ADAM optimizer \cite{Kingma2014Adam:Optimization}, with the default settings and using a constant learning rate of 0.0001. In order to compare the performance of the traditional optimization objective, MAE (\ref{eq:MAE}), and the proposed optimizaton objective, FMAE (\ref{eq:FMAE}),  two DNNs are trained for each auditory model: One DNN using MAE  as the optimization objective, and another DNN using FMAE. Both networks are trained using the same auditory-model parameters $\theta$, with the same choice of seed, network parameters and hyperparameters, the only difference being the optimization objective, resulting in two DNNs with different weights.

\subsection{Verhulst model}
\label{sec:training_verhulst}
The Verhulst auditory model is tested for two different scenarios: 
First, we train DNNs using both the MAE and FMAE, using the same architecture as before, but for input levels between 40 and 120 dB SPL as explained above, using the following model parameters configurations: Normal hearing (Normal), a sloping, mild hearing loss (Slope20\_5) and a flat, mild hearing loss (Flat20). The auditory-model parameters for these configurations can be found online, together with the code for the auditory model \cite{GitHubPublication}. For both scenarios the input-output pairs are extended with 256 samples (13 ms) of temporal context on the left side of the window and 256 (13 ms) on the right side of the window. Additionally, we scale the output by a factor $5\cdot 10^4$, which preserves the relative amplitudes to be consistent with \cite{Baby2021AApplications}. 

Second, we compare our implementation with an already trained network, CoNNear, \cite{Baby2021AApplications}, available at \cite{GitHubHearingTechnology/CoNNear_cochlea}. Note that CoNNear is trained with speech signals all normalized to 70 dB SPL, meanwhile our DNN is trained with speech signals between 40 dB SPL and 120 dB SPL. Since CoNNear and DNN-FMAE have the exact same number of parameters, and the DNN-FMAE has to work for a much larger dynamic range, one would expect the CoNNear to outperform the DNN-FMAE at input levels close to 70 dB SPL, for which the CoNNear is a specialist.

\subsection{Zilany model}
The DNNs trained to emulate the Zilany model use 1024 samples (51 ms) of temporal context on the left side of the window and 256 samples (13 ms) on the right side of the window. We use the N0, N3, N5 and S1 audiogram templates explained in Sec. \ref{sec:method_zilany}.

\section{Results}
\label{sec:results}
In this section we compare our proposed optimization objective, FMAE, to the conventional objective, MAE, using both of the auditory models and their respective auditory model emulators. To do so, we train two deep neural networks that emulate the selected auditory model based on the methodology detailed in Sec. \ref{sec:training}.  The networks are called DNN-MAE and DNN-FMAE, depending on the optimization objectives that were used during training. We assess the performance of the proposed optimization objective, by comparing the performance of the model with evaluation metrics used in state-of-the-art and our own metric, the SER (\ref{eq:SER}). The evaluation of the emulators is performed for both speech, music and pure tones.

\subsection{Response to speech}
The auditory model emulators are trained on a large dataset of speech signals, because accurate modelling of speech is of highest importance in our hearing aid applications, and for the same reason the auditory model emulators are evaluated on speech. A set of 20 random utterances from different talkers that were not used during the training were utilized to obtain the error measurements at each input SPL, leading to a total of 100 test utterances.

\begin{figure*}[htbp]
  \centering
      \includegraphics[scale=0.5]{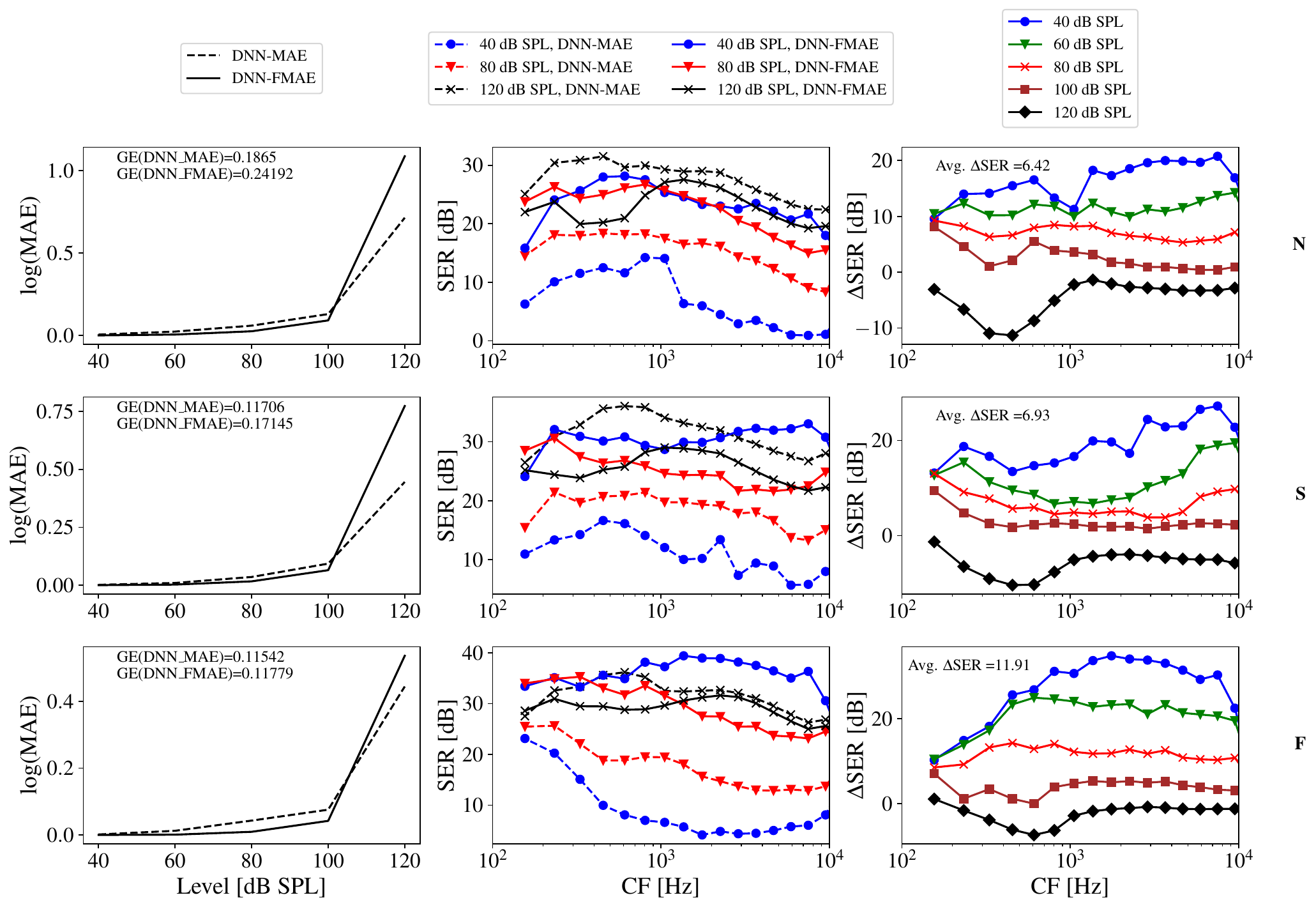}
  \caption{Evaluation metrics computed on speech signals for the Verhulst auditory model in 3 different conditions: N(Normal Hearing), Slope20\_5 (S), Flat20 (F), measured across 20 unseen sentences at each level. CF denotes the characteristic frequency. The CFs of the model have been selected uniformly. DNN-MAE denotes the DNN trained with the MAE as an optimization objective and DNN-FMAE the DNN trained with FMAE as an optimization objective. The first column shows the log(MAE) across different levels, and the global MAE (GE): The average MAE across all levels for the DNN-MAE and DNN-FMAE. The second column shows the error in SER across level for the DNN-MAE and DNN-FMAE. The third column shows the difference in SER between DNN-FMAE and DNN-MAE across different levels, and the average increase in SER. Note the different scaling on the plots.}
  \label{fig:speech-connear}
\end{figure*}

\begin{figure*}[htbp]
  \centering
      \includegraphics[scale=0.5]{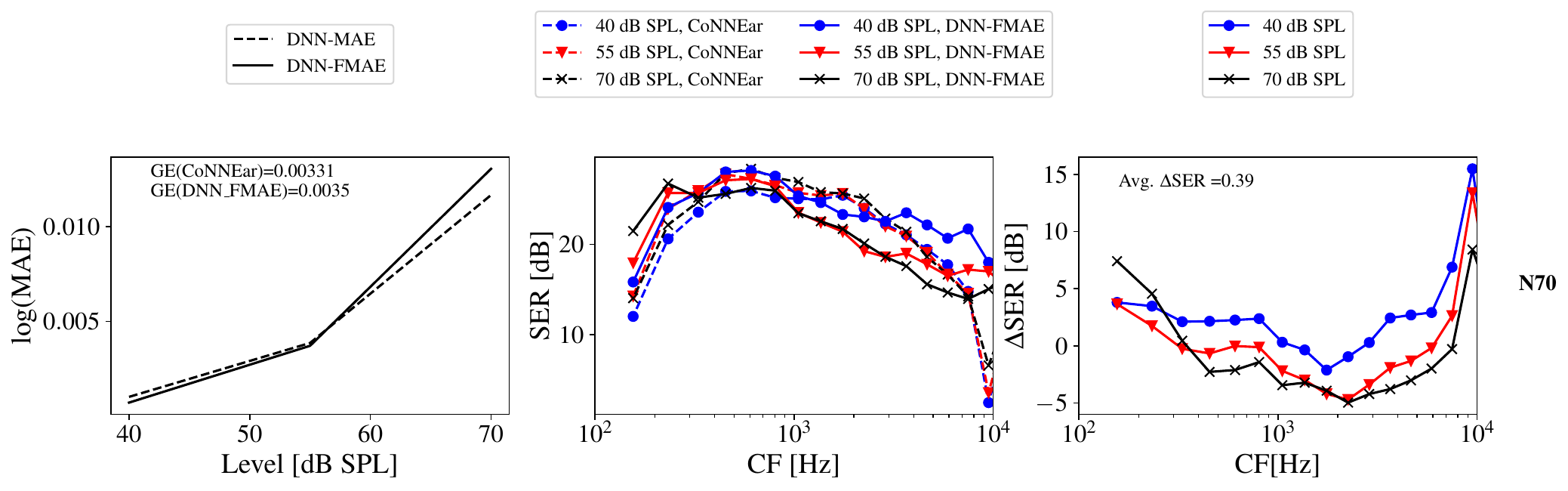}
  \caption{Evaluation metrics computed on speech signals for the Verhulst auditory model for Normal hearing using inputs between 40 and 70 dB SPL (N70), measured across 20 unseen sentences at each level. CF denotes the characteristic frequency. The CFs of the model have been selected uniformly. CoNNear denotes the DNN taken from \cite{Baby2021AApplications}, that was trained at 70 dB SPL, and DNN-FMAE the DNN trained with FMAE as an optimization objective, that was trained with input signals between 40 and 120 dB SPL. The first column shows the log(MAE) across different levels, and the global MAE (GE): The average MAE across all levels for CoNNear and DNN-FMAE. The second column shows the error in SER across level for CoNNear and DNN-FMAE. The third column shows the difference in SER between DNN-FMAE and CoNNear across different levels, and the average increase in SER.}
  \label{fig:speech-verhulst1}
\end{figure*}

\begin{figure*}[htbp]
  \centering
      \includegraphics[scale=0.5]{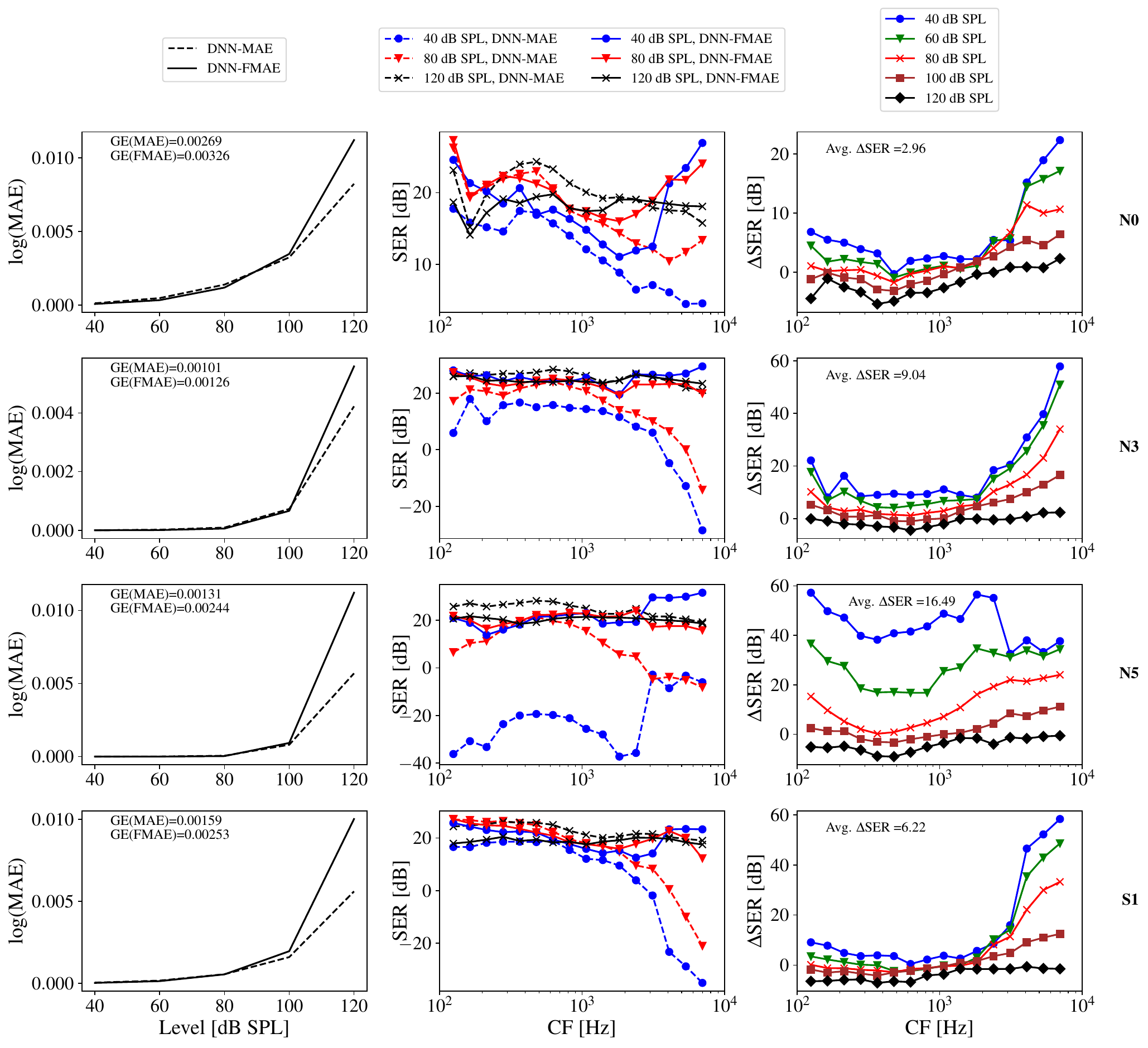}
  \caption{Evaluation metrics computed on speech signals for the Zilany auditory model using 4 different audiograms, N0(Normal Hearing) and 3 types of hearing loss \cite{Bisgaard2010} as measured across 20 unseen sentences. CF denotes the characteristic frequency. The CFs of the model have been selected uniformly. DNN-MAE denotes the model trained with the MAE as an optimization objective and DNN-FMAE the model trained with FMAE as an optimization objective. The first column shows the log(MAE) across different levels, and the global MAE (GE): The average MAE across all levels for the DNN-MAE and DNN-FMAE. The second column shows the error in SER across level for the DNN-MAE and DNN-FMAE. The third column shows the difference in SER between DNN-FMAE and DNN-MAE across different levels, and the average increase in SER. Note the different scaling on the plots.} 
  \label{fig:speech-carney}
\end{figure*}

Figs. \ref{fig:speech-connear}-\ref{fig:speech-carney} show the auditory model emulators response to speech. The first column in all figures shows the logarithm of the Mean Absolute Error, log(MAE), of the two deep neural network emulators, which are computed for each input SPL, similarly to the evaluation metric used in \cite{Nagathil2021ComputationallyProcessing}. Alongside the log(MAE), we compute the global MAE (referred to as GE), representing the average MAE across all input sound levels, as it was used in \cite{Baby2021AApplications}. The second column shows the proposed evaluation criterion, SER (\ref{eq:SER}), calculated at each input SPL. Finally, the third column shows the $\Delta SER$, i.e. the difference in SER between the DNN-FMAE and DNN-MAE models.

Fig. \ref{fig:speech-connear} shows the performance for the Verhulst auditory model emulator for three different auditory model configurations of increasing hearing-loss severity: Normal hearing (N), sloping hearing loss (S) and flat hearing loss (F). Fig. \ref{fig:speech-connear}, Col. 1 demonstrates that the auditory model emulators trained using a conventional optimization objective, MAE, results in emulators that are dominated by high input levels: For high input levels, the log(MAE) is lower for the DNN-MAE, than the DNN-FMAE, but as the input SPL decreases, the log(MAE) of the DNN-FMAE is lower than the DNN-MAE. This is of critical importance, as the auditory model emulators need to function across an ecologically valid range of input SPL for developing HLC, not only the extremely high SPL inputs. However, the MAE metric is difficult to interpret, primarily for two reasons: 1) The magnitude of the error can be seen to vary across both input SPL, hearing loss and auditory models (see Fig. \ref{fig:speech-carney}, Col. 1), which makes comparison across auditory models and parameter configurations difficult, and 2) conventionally, the MAE metric is averaged across frequency channels, where the energy distribution might be highly skewed, making it impossible to assess the performance of individual frequency channels. Note, that even if the MAE is calculated for each level and frequency channel, it would still suffer from problem 1).  The second column shows our proposed metric, the SER, which is both dependent on input SPL and frequency channel. Here, we observe that the SER of the DNN-MAE varies a lot across input SPL and frequency channel, meanwhile the SER of the DNN-FMAE tends to vary much less, leading to a more consistent performance. In the first row, the SER of the 40 dB SPL inputs show a large dip at the lowest CFs for both the DNN-MAE and the DNN-FMAE,  due to the RF of the emulators being too short compared to the CF where the impulse response of the Verhulst auditory model is longest, cf. Sec. \ref{sec:DNN_RF}. In the last column, we observe that the DNN-FMAE achieves a monotonically increasing benefit as the input SPL decreases, and also an increase in average $\Delta$SER as hearing loss severity increases. 

\begin{figure}[!h]
  \centering
      \includegraphics[scale=0.5]{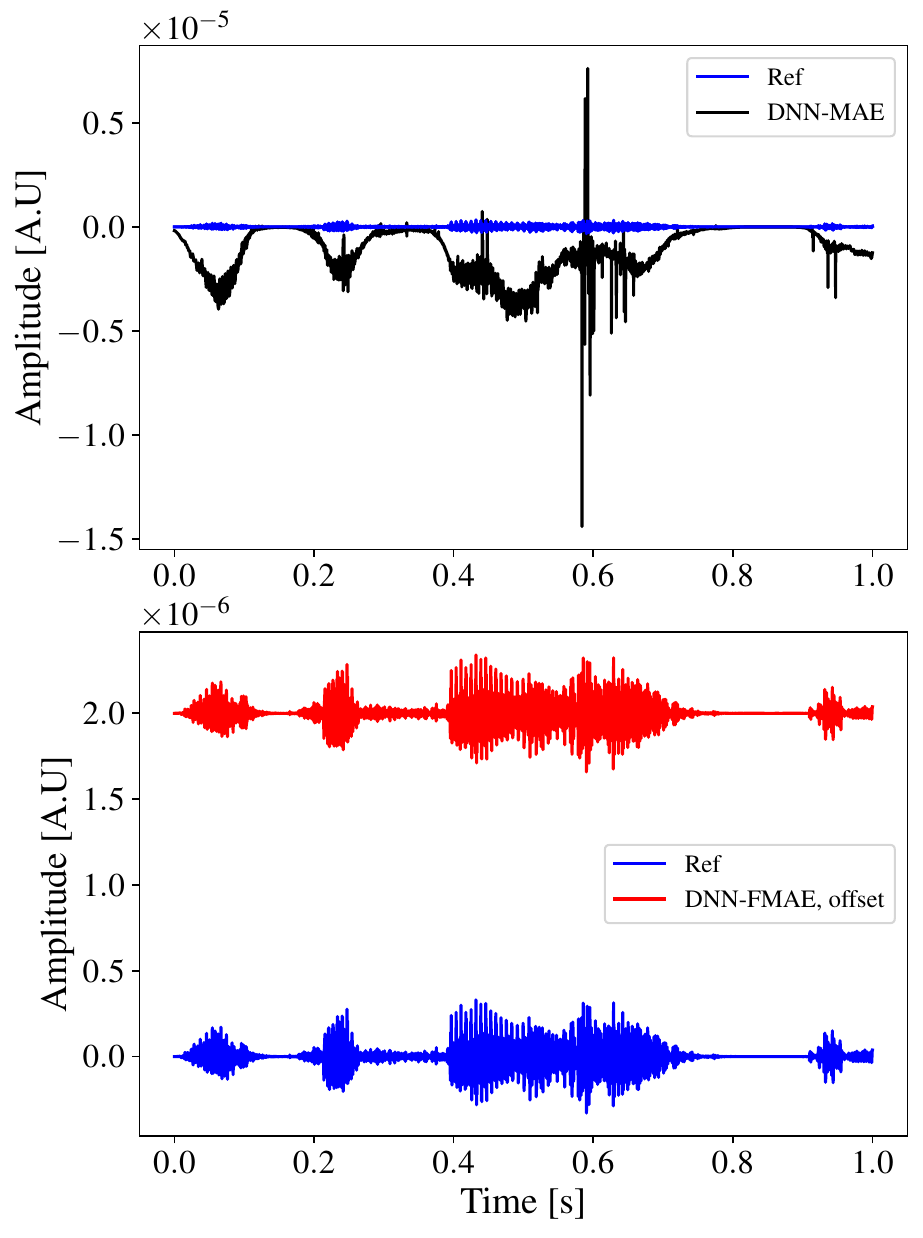}
  \caption{Sample output of the Zilany auditory model at CF = 8 kHz, parameterized by a N3 audiogram. The input is a segment of a speech signal segment normalized to 40 dB SPL. Ref denotes the ground truth auditory model, DNN-MAE the DNN auditory model emulator trained using MAE and DNN-FMAE, the DNN auditory emulator trained using FMAE. An offset of $+2 \times 10^{-6}$ has been added to the output of the DNN-FMAE  to avoid complete overlap with the reference.}
  \label{fig:inner_speech}
\end{figure}

Fig. \ref{fig:speech-verhulst1} shows the performance metrics for the Verhulst auditory model emulator, where we compare the DNN-FMAE to the CoNNear. Note that, although the DNNs have the same architecture, there might be differences that can be attributed to differences in the training, such as neural network weight initialization, and the training set itself, all leading to different training landscapes, that might cause faster or better training. Therefore, in this case the focus should be on the relative improvement in performance between high and low energy regions, not the absolute improvement. The last two columns show similar results as before, namely that the largest modelling  performance increase, in terms of SER,  is in the low-energy regions. However, the middle column also shows that the high frequency CFs of the CoNNear shows very poor performance, almost 0 SER, regardless of input SPL. This finding is consistent with \cite{Baby2021AApplications}, where the authors note that the CoNNear does not perform well for input frequencies above 8 kHz. This result shows that even though the DNN-FMAE is trained to cover a much larger dynamic range, up to 120 dB SPL, it performs better in the low-energy regions. Additionally, for the 40 dB SPL input we also see a large dip at low frequencies, which again can be explained by the RF being too small.
Fig. \ref{fig:speech-carney} shows the performance metrics for the Zilany auditory model emulator for four different auditory model configurations: N0 (Normal hearing), N3 (moderate hearing loss), N5 (severe hearing loss), and S1 (steep hearing loss), explained in detail in Sec. \ref{sec:method_zilany}. In general, we make similar observations as for the Verhulst auditory model. In the first column the DNN-MAE always achieves a lower GE than the DNN-FMAE. Inspecting the first row (N0) and the second and third column , there is a dip in performance at approximately 1 kHz for the low input levels: The DNN-FMAE performs similarly to the DNN-MAE, indicating that although FMAE provides a balanced optimization objective, it does not address performance issues, which could stem from a sub-optimal choice of DNN architecture or sub-optimal training of the DNN. However, when  comparing the first row (N0) to the second (N3), we observe that the DNN-FMAE outperforms the DNN-MAE  by a large margin. For high frequency CFs, the 40 dB SPL input results in an average SER of approximately -30 dB for the DNN-MAE, and to put this into perspective, we plot in Fig. \ref{fig:inner_speech} a segment of a 40 dB SPL speech signal for the 8 kHz CF. From Fig. \ref{fig:inner_speech} it is evident that the DNN-MAE produces a very poor approximation, while the DNN-FMAE closely resembles the reference auditory model. The last row in Fig. \ref{fig:speech-carney} corresponds to a steeply sloping hearing loss (S1), where the major difference between the DNN-MAE and DNN-FMAE is restricted to high frequencies. This difference can be attributed to the steep profile of the S1 audiogram.  For both the Verhulst model and the Zilany model, the benefit of the proposed method becomes increasingly evident as the hearing loss is increased. In general, we observe that the MAE is an inadequate metric for model choice, and these results, combined with the results from the Verhulst auditory model, show that the FMAE optimized emulators perform significantly better than the MAE optimized emulators, except for the very extreme inputs of 120 dB SPL, but that is a reasonable trade-off. The inadequacy of the MAE for auditory-model emulation is particularly illustrated in the case of the N5 audiogram where the global MAE is twice as large for the DNN-FMAE compared to the DNN-MAE, but the average $\Delta\text{SER}$ is 16.49. Thus, an  optimization objective that takes into account the energy distribution of the training set, given by the input signals, the respective auditiory model and parameter configuration, helps to achieve an auditory model emulator that scales across both input SPL, the CFs of the auditory model, choice of auditory model and hearing-loss profiles. 

\subsection{Response to music}

\begin{figure*}[htbp]
  \centering      \includegraphics[scale=0.53]{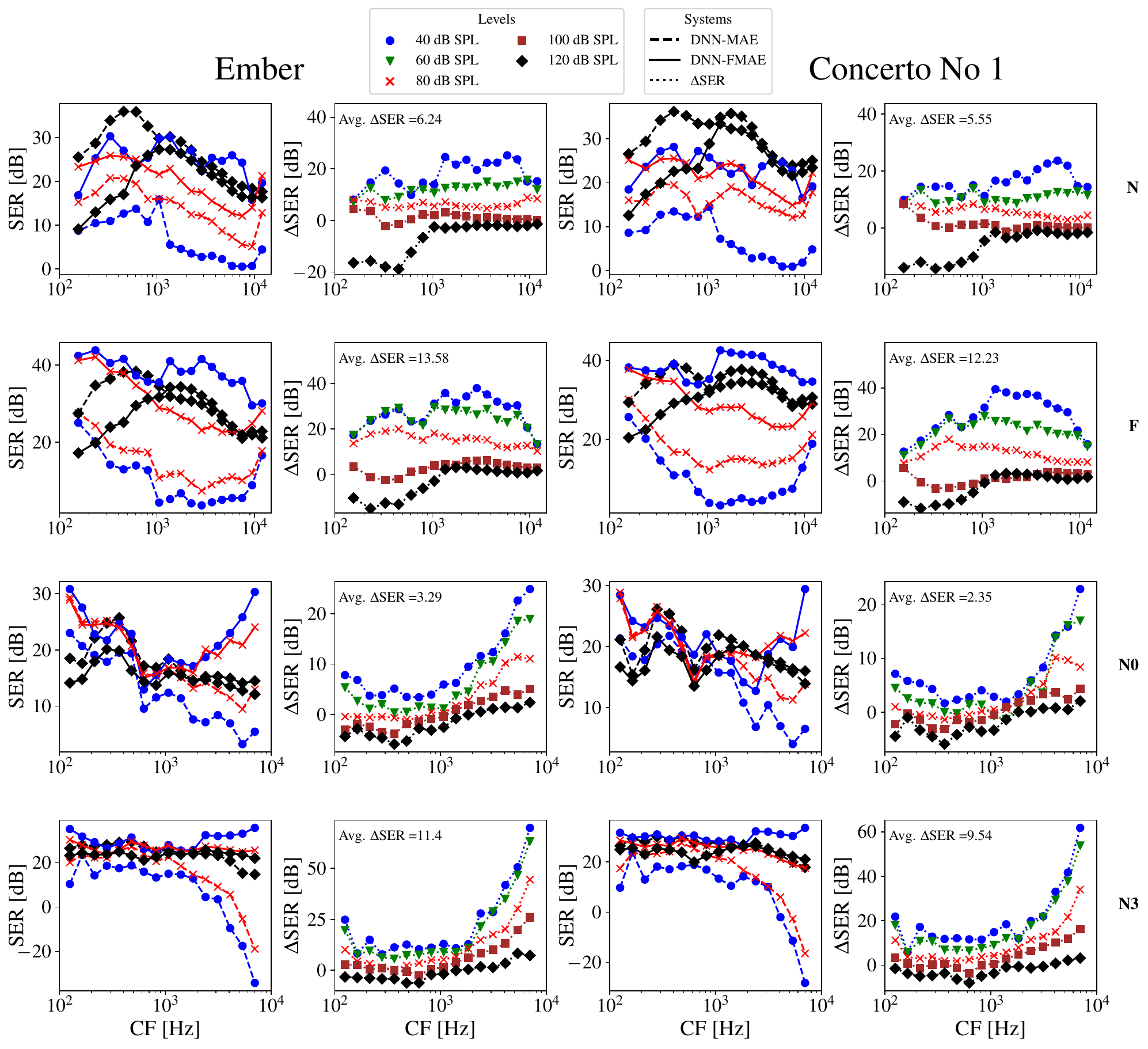}
  \caption{Evaluation metrics computed on two musical recordings: Ember by Kemaca Kinetic  \cite{KineticEmberYouTube}, shown in the first two columns  and Piano Concerto No 1 by Tchaikovsky \cite{TchaikovskyTchaikovskyYouTube}, shown in the last two columns. The first two rows show the Verhulst model, parameterized with Normal hearing (N) and Flat\_20(F) and the last two rows shows the Zilany model, parameterized with Normal hearing (N0) and the N3 audiogram \cite{Bisgaard2010}. CF denotes the characteristic frequency. The CFs of the model have been selected uniformly. DNN-MAE denotes the model trained with the MAE as an optimization objective and DNN-FMAE the model trained with FMAE as an optimization objective. The first and third columns shows, for Ember and Concerto No 1, respectively, shows the error in SER across level for the DNN-MAE (dashed lines) and DNN-FMAE (solid lines). The second and fourth column, for Ember and Concerto No 1, respectively, shows the difference in SER between DNN-FMAE and DNN-MAE across different levels, and the average increase in SER. Note the different scaling on the plots.}
  \label{fig:music}
\end{figure*}

To verify that our proposed optimization objective generalizes to unseen acoustic material, we test our emulators - that were trained on only speech signals as described in Sec. \ref{sec:training} - on two musical pieces: 
\begin{enumerate}
    \item A rock-jazz recording, Ember, by Kemaca Kinetic, featuring electric guitar, electric bass and drumset \cite{KineticEmberYouTube}. We use the first 30 seconds of the recording.
 \item A classical piece, Concerto No. 1 in B-Flat Minor, op 23, written by Pjotr Tjajkovskij and played by Van Cliburn. The piece features a piano and a symphonic orchestra \cite{TchaikovskyTchaikovskyYouTube}. We use the first 30 seconds of the recording.
\end{enumerate}
For the sake of brevity, we conduct this experiment for two different  model parameterizations, for each auditory model: Normal (N) and Flat20 (F) for the Verhulst model and  N0, N3 for the Zilany model. Additionally, we discard the MAE measure, as in the previous experiment it was found to not provide much useful information, as compared to the SER. The results of the experiment is shown in Fig. \ref{fig:music}.
We find that the DNN-FMAE performs relatively better than DNN-MAE, as measured across different CFs and levels, and that the $\Delta \textrm{SER}$ measure mirrors the results found in the speech experiment, except for the Verhulst model at low frequencies and at the highest SPL. Upon inspecting the activation functions of the DNN-FMAE, we find that the performance decrease at high input levels and low CFs is because the Tanh activation saturates. The saturation can be explained by the training dataset (speech) having a relatively low peak-to-average power ratio (PAPR) at the low CFs compared to music, leaving not enough headroom at the highest SPL. The saturation can be mitigated by including larger amplitudes in the training dataset, e.g. by including higher input SPLs in the training dataset. Thus, the results are consistent with our previous findings and indicates that our method also generalizes to out-of-distribution acoustic material.

\subsection{Response to pure tones}
To evaluate the generalization capability of auditory model emulators, state-of-the-art literature uses pure-tone stimuli, as this type of stimuli was used to verify the original auditory models from physiological data. The root mean square (RMS) value of the pure-tone response is calculated for each CF, and the patterns that emerge from this process are referred to as excitation patterns \cite{Baby2021AApplications}. The excitation patterns of the emulators are compared to those of the reference auditory model for a set of different pure-tone frequencies at different input SPLs. We repeat these experiments, and compute the resulting excitation patterns using emulators trained with the conventional MAE optimiziation objective and our proposed optimization objective, the FMAE. 

\begin{figure*}[htbp]
  \centering
      \includegraphics[scale=0.53]{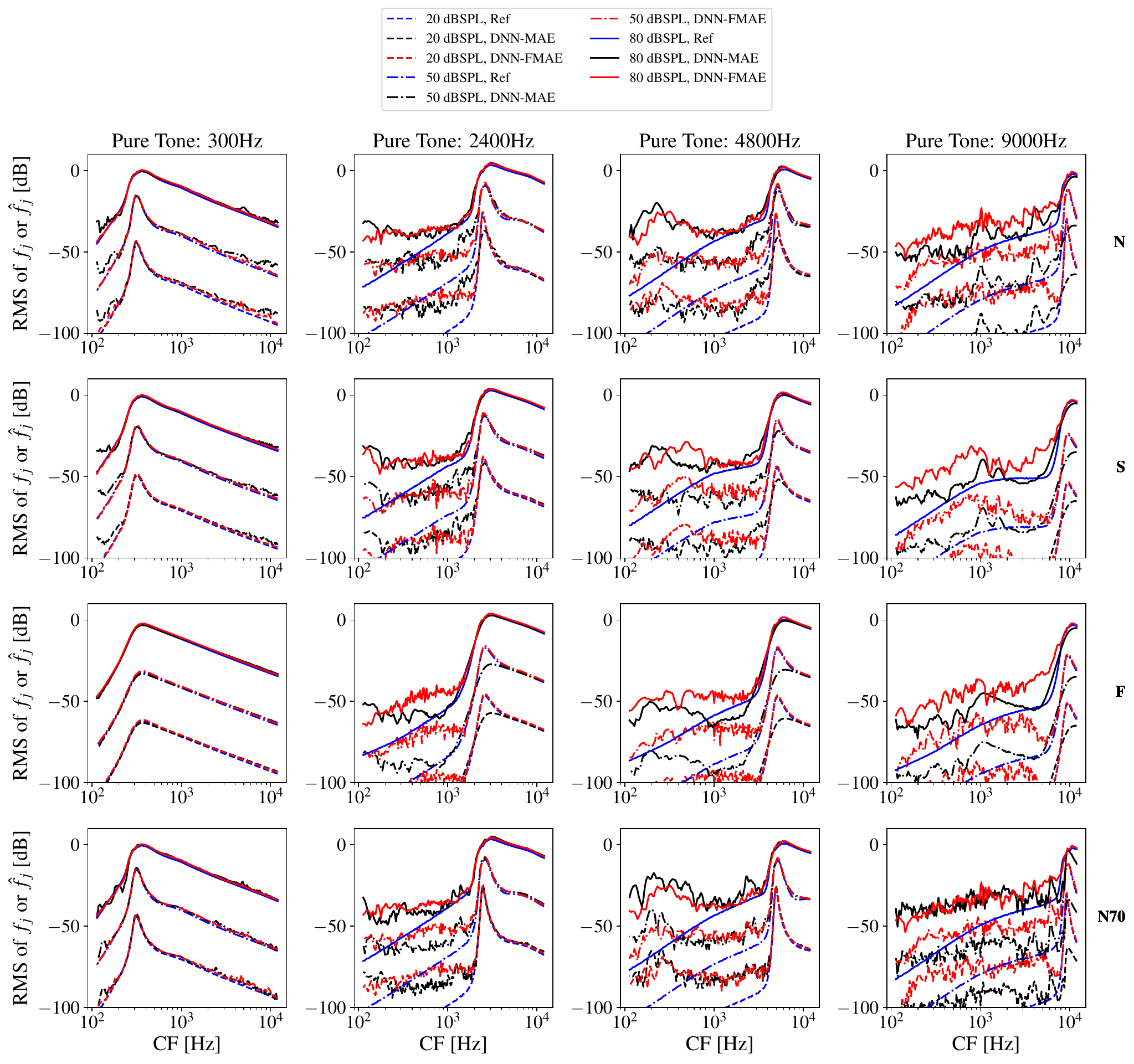}
  \caption{Tonal response to pure tones at different sound input levels and frequencies for 4 different conditions for the Verhulst model: N(Normal Hearing), Slope20\_5 (S), Flat20 (F) and (N70), a DNN trained with input SPLs between 40 and 70 dB compared with CoNNear from \cite{Baby2021AApplications}. CF denotes the characteristic frequency. The CFs of the model have been selected uniformly. DNN-MAE denotes the DNN trained with the conventional MAE as an optimization objective (CoNNear in the case of N70) and DNN-FMAE the model trained with FMAE as an optimization objective.}
  \label{fig:tonal_verhulst}
\end{figure*}

Fig. \ref{fig:tonal_verhulst} shows the excitation patterns for the Verhulst auditory model and the emulators in four different scenarios: Normal hearing (N), sloping hearing loss (S), flat hearing loss (F) and a comparison with CoNNear (N70), explained in Sec. \ref{sec:training_verhulst}. In row 1, we observe that both the DNN-MAE and DNN-FMAE capture the correct behaviour for lower frequencies, while the DNN-MAE fails as the pure-tone frequency increases, i.e. at a pure-tone frequency of 4800 Hz. The same pattern can be seen in the second row (S). As the hearing-loss severity  is increased in row 3 (F), the DNN-MAE starts to fail at a pure-tone frequency of 2400 Hz, and shows a larger discrepancy compared to the reference at a pure-tone frequency of 4800 Hz. In general, the DNN-MAE captures the peak of the excitation patterns at high levels, while the peak is not captured at lower levels since the energy of the lower levels are negligible in terms of MAE. In the last case (N70), we compare the DNN-FMAE to the pre-trained CoNNear. Here we find that the CoNNear performs comparably to the DNN-FMAE for all input frequencies, except 9 kHz, where the CoNNear performs poorly. This finding is consistent with the results from Fig. \ref{fig:speech-verhulst1}, where we observe very poor performance for very high frequency CFs for the CoNNear. Our proposed DNN-FMAE, using the exact same architecture and similar training data, shows good performance across input SPL, frequency channels and hearing-loss severity.

\begin{figure*}[htbp]
  \centering
      \includegraphics[scale=0.53]{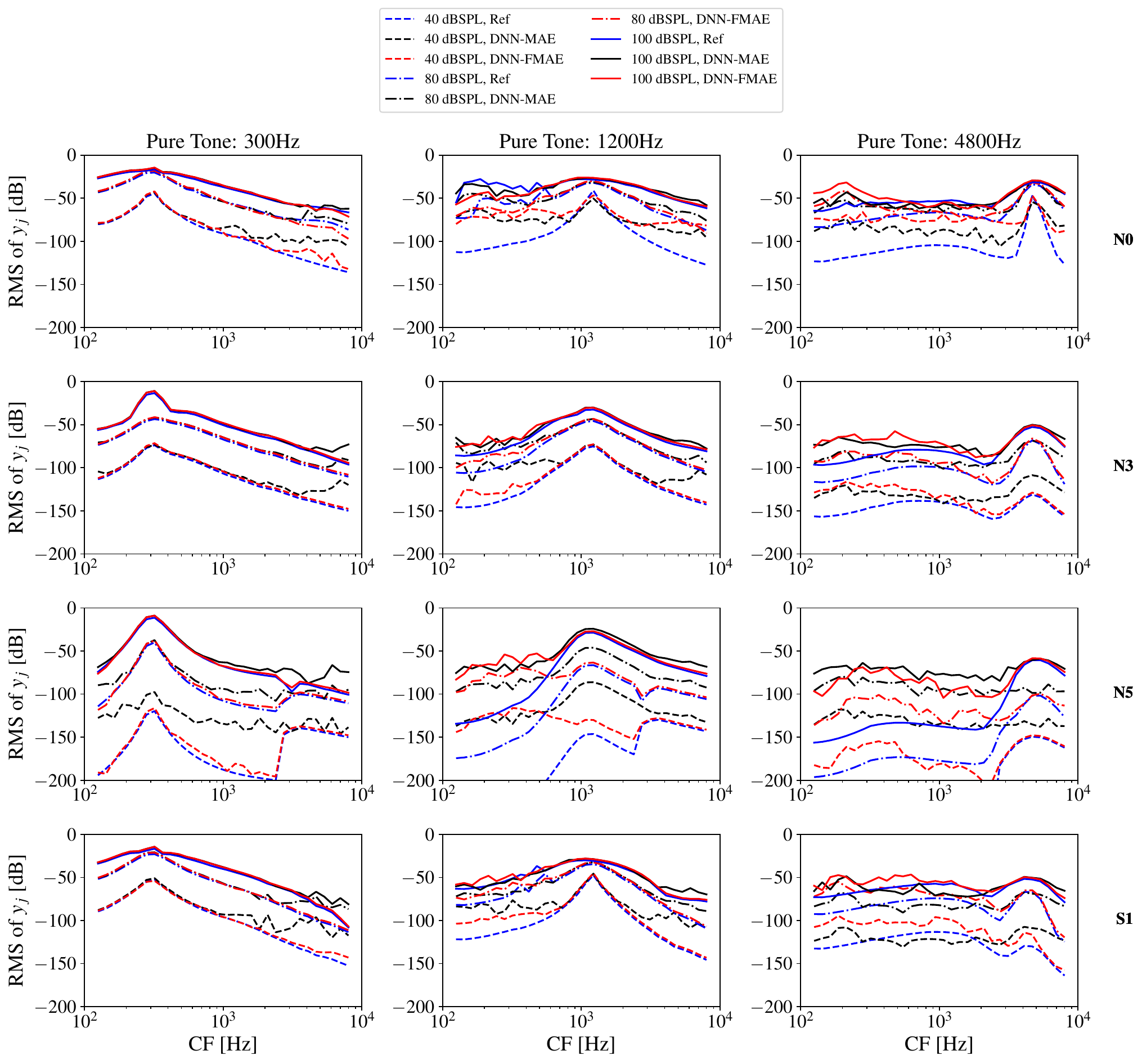}
  \caption{Tonal response to pure tones at different sound input levels and frequencies for 4 different conditions for the Zilany model: N0(Normal Hearing) and 3 types of hearing loss \cite{Bisgaard2010} in varying degree. CF denotes the characteristic frequency. The CFs of the model have been selected uniformly. DNN-MAE denotes a DNN trained with the conventional MAE as an optimization objective and DNN-FMAE the model trained with FMAE as an optimization objective.}
  \label{fig:tonal_carney}
\end{figure*}

Fig. \ref{fig:tonal_carney} shows the excitation patterns for the Zilany Auditory model for 4 different scenarios: N0 (Normal hearing), N3 (moderate hearing loss), N5 (severe hearing loss), and S1 (steep hearing loss), explained in detail in Sec. \ref{sec:method_zilany}. In the first row (N0), both the DNN-MAE and DNN-FMAE align well with the reference for the lowest frequency input. However, as the pure-tone frequency increases both models do not track with the reference, particularly at CFs lower than the pure-tone frequency. This pattern is repeated in both row 2 (N3) and row 3 (N5), where the DNN-FMAE aligns much better with the reference at CFs close to the pure-tone frequency at lower input SPLs, meanwhile the DNN-MAE is far from the reference. In row 4 (S1), both the DNN-MAE and DNN-FMAE align well with the reference at low pure-tone frequencies, but align poorly with the reference at high pure-tone frequency. The general pattern is that the models perform worse as the pure-tone frequency is increased, irrespective of hearing loss. By inspecting the time-domain representation of the auditory model output, we find that the worsening of performance is due to the auditory model behaving differently for low and mid-to-high frequency input. For low-frequency inputs, all channels display similar behavior and produce a filtered sinusoidal output, while for mid- and high-frequency inputs, the auditory model produces a click response at lower CFs and behaves as an envelope detector at mid-to-high CFs. Both the DNN-MAE and the DNN-FMAE are unable to capture this behaviour, presumably because only broadband signals were used during training, and this kind of behaviour was never seen by the emulator. The Verhulst auditory model on the contrary displays uniform behaviour: All channels are to first order identical if one normalizes out the quality-factor and the CF, and regardless of whether a low or high-frequency pure-tone stimulus is used, the resulting output will always be a filtered sinusoidal waveform.

\onecolumn
\twocolumn

\section{Discussion}
The proposed optimization objective, the FMAE (\ref{eq:FMAE}), depends directly on the auditory model, the hearing loss and the input signals, allowing the trained AME to function across a wide range of input levels and hearing losses. Instead of directly changing the loss function, it was recently suggested to transform the inner representation, using a log-based transformation that depends on a compression parameter, $d$ \cite{Nagathil2023WaveNet-basedModel}. However, in \cite{Nagathil2023WaveNet-basedModel} the log-based transformation was tested on the Zilany model, using only the normal hearing parameterization. It is therefore still unclear if the log-based transformation generalizes to different hearing losses and auditory models. Assuming the method in \cite{Nagathil2023WaveNet-basedModel} generalizes, there could be one disadvantage of such a method: Since the energy distribution of the inner representation changes drastically for different auditory model and hearing loss, one could imagine that, for each combination of auditory model and hearing loss, one would need to choose an optimal $d$. Our proposed optimization objective is hyperparameter-free and thus avoids this problem all-together, since the parameters can be easily derived from the training data itself.

In Sec \ref{sec:activations}, we revisited  arguments made in \cite{Baby2021AApplications} and \cite{Drakopoulos2021ASynapses} on why some activation functions are suitable for auditory modelling. Particularly, for the CoNNear it was argued that the hyperbolic tangent (Tanh) resembles the input-output relations of the outer hair cells, and thus is suitable as an activation function. The results from Figure \ref{fig:speech-connear}, \ref{fig:speech-verhulst1} and \ref{fig:tonal_verhulst} suggest that there might be additional reasons: If one trains a CoNNear using inputs between 40 and 120 dB SPL, using the MAE, modelling performance is significantly worse at lower input levels and higher CFs. These results imply that if the CoNNear is trained at input levels where the energy distribution of the inner representation is compressed, the Tanh activation functions of the DNN induces an implicit weighting of lower level inputs. To see this, consider the following: 1) During backpropagation, the weight updates are proportional to the derivative of the activation function to which it connects\cite{Bishop2006}, and 2) the derivative of Tanh is expansive, i.e. a large derivative for small activations. Thus, the contributions to the weight-update for lower level inputs - leading to lower output magnitudes and activations - are given a larger emphasis by the derivative of the activation function, to some degree counteracting the smaller activations caused by the lower level inputs. Thus a Tanh activation allows the model to generalize, to some extent, downwards in level.

While auditory models, in general, provide rich and accurate descriptions of the auditory pathway, they still might not properly model aspects of physiological behaviour that might be relevant to perception, e.g. efferent control of hair cells in the cochlea. There are several proofs showing that DNNs are universal function approximators if they satisfy a number of conditions, such as having sufficiently many neurons \cite{Hornik1989MultilayerApproximators}. In the best case, assuming the parameters for such a DNN can be found, the DNN will inherit the limitations of the original auditory models, and in worst case the DNN might not capture relevant perceptual features of the original auditory model. Thus, the applicability of DNN-based auditory models to signal processing is directly limited by the original auditory models ability to model perceptually relevant phenomena. In particular, we found that the DNNs might not accureately represent artificial signals, souch as sinusoids. We can imagine one specific scenario where the ability to represent sinusoid might be important: Some DNNs for noise reduction are based on an encoder-decoder structure, where the decoder can introduce strong aliasing artifacts. If the DNN is optimized using an auditory model emulator, which does not represent, and therefore properly penalize these components, the resulting noise reduction strategy might contain sinusoidal components.

\section{Conclusion}
%\vspace{-150pt}
\label{sec:conclusion}
We have demonstrated that in order to effectively train deep neural networks to simulate auditory models capable of replicating varying degrees of hearing loss, it is crucial to take into account the energy distributions of the auditory models, which are both level- and frequency-dependent, along with the signals that are typically applied to them. We have proposed a straightforward yet powerful optimization objective, Frequency-and-level-dependent Mean Absolute Error, that modifies the conventional optimization objective, by taking into account the energy distributions in the inner representations of the auditory models. This has resulted in significant improvements in the modeling performance across auditory models, their frequency channels and input sound level, when compared to the conventional machine learning optimization objective used in existing auditory model deep-neural-network emulation schemes. Furthermore, we propose to use a normalized evaluation criteria, the signal-to-error ratio, that is easier to interpret, and readily shows the level-and-frequency-dependent performance of the auditory-model emulators. Our findings demonstrate that not only are conventional optimization objectives poor choices of optimization objectives: They are also poor metrics to employ for model selection. We, therefore, recommend that future modeling efforts avoid the use of conventional machine-learning objectives as optimization objectives or evaluation metrics in a straightforward manner.

 For profound and severe hearing losses, the loudness growth function is steep resulting in a large dynamic range in the inner representation space of the auditory model. This causes $\alpha_{j,l}$ to increase rapidly as $l$ decreases. Thus, interpolating in the log-domain, rather than using a nearest-neighbour approach, helps to stabilize the convergence of the optimization objective. If the DNN is only trained on the discrete levels in $L$, $a_j(x)$ will be identical to $\alpha_{j,l}$.

\bibliographystyle{IEEEtran}
\bibliography{IEEEabrv,references}

\end{document}